\documentclass[twocolumn]{aastex6}
\usepackage{graphicx}
\usepackage{amssymb,amsfonts,amsmath,amstext,amsgen,amsopn,amsxtra,indentfirst,times}

\begin{document}

\title{Retrieval Analysis of the Emission Spectrum of WASP-12b: \\ Sensitivity of Outcomes to Prior Assumptions and Implications for Formation History}

\author{Maria Oreshenko\altaffilmark{1}}
\author{Baptiste Lavie\altaffilmark{1,2}}
\author{Simon L. Grimm\altaffilmark{1}}
\author{Shang-Min Tsai\altaffilmark{1}}
\author{Matej Malik\altaffilmark{1}}
\author{Brice-Olivier Demory\altaffilmark{1}}
\author{Christoph Mordasini\altaffilmark{1}}
\author{Yann Alibert\altaffilmark{1}}
\author{Willy Benz\altaffilmark{1}}
\author{Sascha P. Quanz\altaffilmark{3}}
\author{Roberto Trotta\altaffilmark{4}}
\author{Kevin Heng\altaffilmark{1}}
\altaffiltext{1}{University of Bern, Center for Space and Habitability, Gesellschaftsstrasse 6, CH-3012, Bern, Switzerland \\ Emails: maria.oreshenko@csh.unibe.ch, kevin.heng@csh.unibe.ch}
\altaffiltext{2}{Observatoire de l'Universit\'e de Gen\`eve 51 chemin des Mailettes, 1290, Sauverny, Switzerland}
\altaffiltext{3}{ETH Z\"{u}rich, Department of Physics, Wolfgang-Pauli-Strasse 27, CH-8093 Z\"{u}rich, Switzerland}
\altaffiltext{4}{Astrophysics Group, Imperial College London, Blackett Laboratory, Prince Consort Road, London SW7 2AZ, U.K.}

\begin{abstract}
We analyze the emission spectrum of the hot Jupiter WASP-12b using our \texttt{HELIOS-R} retrieval code and \texttt{HELIOS-K} opacity calculator.  When interpreting Hubble and Spitzer data, the retrieval outcomes are found to be prior-dominated.  When the prior distributions of the molecular abundances are assumed to be log-uniform, the volume mixing ratio of HCN is found to be implausibly high.  A \texttt{VULCAN} chemical kinetics model of WASP-12b suggests that chemical equilibrium is a reasonable assumption even when atmospheric mixing is implausibly rigorous.  Guided by (exo)planet formation theory, we set Gaussian priors on the elemental abundances of carbon, oxygen and nitrogen with the Gaussian peaks being centered on the measured C/H, O/H and N/H values of the star.  By enforcing chemical equilibrium, we find substellar O/H and stellar to slightly superstellar C/H for the dayside atmosphere of WASP-12b.  The superstellar carbon-to-oxygen ratio is just above unity, regardless of whether clouds are included in the retrieval analysis, consistent with \cite{madhu11}.  Furthermore, whether a temperature inversion exists in the atmosphere depends on one's assumption for the Gaussian width of the priors.  Our retrieved posterior distributions are consistent with the formation of WASP-12b in a solar-composition protoplanetary disk, beyond the water iceline, via gravitational instability or pebble accretion (without core erosion) and migration inwards to its present orbital location via a disk-free mechanism, and are inconsistent with both in-situ formation and core accretion with disk migration, as predicted by \cite{madhu17}.  We predict that the interpretation of James Webb Space Telescope WASP-12b data will not be prior-dominated.
\end{abstract}

\keywords{planets and satellites: atmospheres}

\section{Introduction}

\begin{figure}
\vspace{-0.1in}
\begin{center}
\includegraphics[width=\columnwidth]{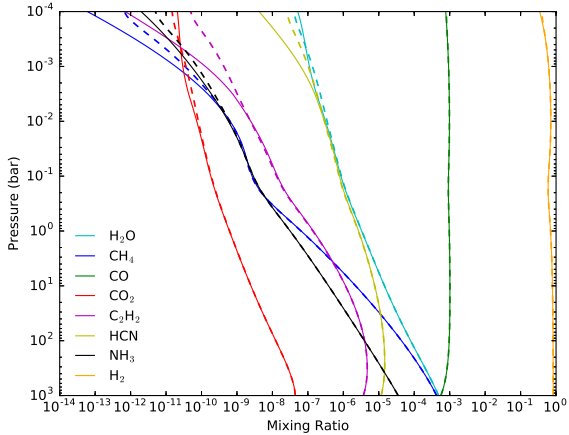}
\end{center}
\vspace{-0.1in}
\caption{Chemical kinetics model of WASP-12b computed using the open-source \texttt{VULCAN} code \citep{tsai17} and adopting the eddy mixing coefficient to be $K_{\rm zz}= 10^{12}$ cm$^2$ s$^{-1}$ (dashed curves).  The solid curves are the molecular abundances in chemical equilibrium.  Photochemistry has been omitted, as it is subdominant due to the high temperatures involved.  Note that we do not use chemical-equilibrium boundary conditions at the bottom of the model atmosphere, but rather zero-flux boundary conditions.  The temperature-pressure profile used is taken from the retrieval model with equilibrium chemistry and no clouds with prior assumptions on the elemental abundances set to twice the measurement errors (``EB,$\times 2$"; see text for details).  The carbon-to-oxygen ratio is set to unity.  Emission spectra typically probe $\sim 0.01$--1 bar, which implies that chemical equilibrium is a good assumption for the atmosphere of WASP-12b.}
\label{fig:chemistry}
\end{figure}

WASP-12b is a well-studied hot Jupiter that has generated ample debate and controversy in the published literature.  With an equilibrium temperature in excess of 2500 K \citep{hebb09}, it serves as a high-temperature laboratory for the study of atmospheric chemistry.  We expect equilibrium chemistry to be a reasonable approximation, as the high temperatures should overwhelm disequilibrium due to atmospheric circulation or photochemistry.  Figure \ref{fig:chemistry} shows a chemical kinetics model of WASP-12b computed using our open-source \texttt{VULCAN} code \citep{tsai17}, which lends support to this expectation.  Even with an eddy mixing coefficient of $K_{\rm zz} \sim 10^{12}$ cm$^2$ s$^{-1}$, the model atmosphere is well-described by chemical equilibrium.\footnote{Using a sound speed of $c_s \sim 1$ km s$^{-1}$ and a pressure scale height of $H \sim 100$ km yields $K_{\rm zz} \sim c_s H \sim 10^{12}$ cm$^2$ s$^{-1}$.  This may be considered an upper limit as vertical flow velocities are typically subsonic.}  Later in the study, we will demonstrate that enforcing chemical equilibrium as a prior assumption circumvents the debate over whether the inferred molecular abundances in WASP-12b are physically and chemically plausible \citep{madhu12,stevenson14b,hl16}.

An active topic of interest associated with WASP-12b is the inferred carbon-to-oxygen (C/O) ratio of its atmosphere, starting with the claim of \cite{madhu11} and \cite{madhu12} that it equals or exceeds unity based on analyzing its emission spectrum.  This inference on the C/O, if true, would imply interesting constraints on the formation and/or evolutionary history of the exoplanet \citep{oberg11,ad14,madhu14b,mordasini16,oberg16,ad17,brew17,esp17,madhu17}, as the C/O of its star has been measured to be $0.48 \pm 0.08$ \citep{teske14}.  In fact, when compared to a sample of exoplanet-bearing stars, WASP-12 is unremarkably Sun-like \citep{teske14,bf16}.  \cite{line14} inferred C/O$=0.51$ from their retrieval analysis, but their inferred volume mixing ratio for CO$_2$ was nearly 0.06, a factor of 26 higher than that for CO, which is chemically implausible unless the metallicity is several orders of magnitude above solar \citep{madhu12,hl16}.  \cite{stevenson14b} performed a uniform analysis of Hubble and Spitzer secondary-eclipse data, subjected them to a retrieval analysis and found a bimodal distribution for C/O.  Oxygen-rich models were ruled out on the basis of chemical implausibility.  By contrast, \cite{kreidberg15} ruled out a carbon-rich interpretation from analyzing the transmission spectrum of WASP-12b.

These properties of WASP-12b, and the attention it has garnered in the community, compel us to perform our own retrieval analysis of its emission spectrum, which probes the dayside of the exoplanet.  Although no new data is being analyzed in the present study, we add value by offering an independent analysis using our own suite of tools \citep{fortney16}.  Furthermore, we  use updated and previously unavailable and/or unused opacities for H$_2$O and CH$_4$.  The high-temperature water line lists were published by \cite{barber06}, while the high-temperature CH$_4$ line lists were published by \cite{y13} and \cite{yt14}.  For example, \cite{line14} did not include HCN opacities in their retrievals and used non-\texttt{ExoMol} CH$_4$ and H$_2$O opacities.  The studies of \cite{madhu11} and \cite{madhu12} also did not use the \texttt{ExoMol} CH$_4$ and H$_2$O opacities.

In the current study, our focus is on elucidating the dependence of the retrieval outcomes on the prior assumptions set on the metallicity or mixing ratios (relative molecular abundances by number).  By ``metallicity", we specifically mean the elemental abundances of carbon (C/H), oxygen (O/H) and nitrogen (N/H), since our 6-molecule analysis only includes the major carbon-, oxygen- and nitrogen-bearing species in their gaseous form.  The assumptions made on the prior distributions of input parameters is an issue that has not been treated in detail in the literature.  Log-uniform prior distributions are often assumed (sometimes without explicitly being stated), based on the misconception that they are the most plausible assumption---erroneously termed ``uninformative priors" or ``uninformed priors"---in the absence of further evidence \citep{trotta08}.  The key finding of our study is that conclusions, based on analyzing currently available data, drawn on C/O and chemistry are strongly tied to our prior assumptions, which are in turn informed by our ideas of physics and chemistry.  Given assumptions on the priors, we then interpret the outcomes, using published studies of (exo)planet formation, by assuming that the retrieved elemental abundances are representative of the bulk composition of the exoplanet.

\section{Methodology}

\begin{figure}
\vspace{-0.1in}
\begin{center}
\includegraphics[width=\columnwidth]{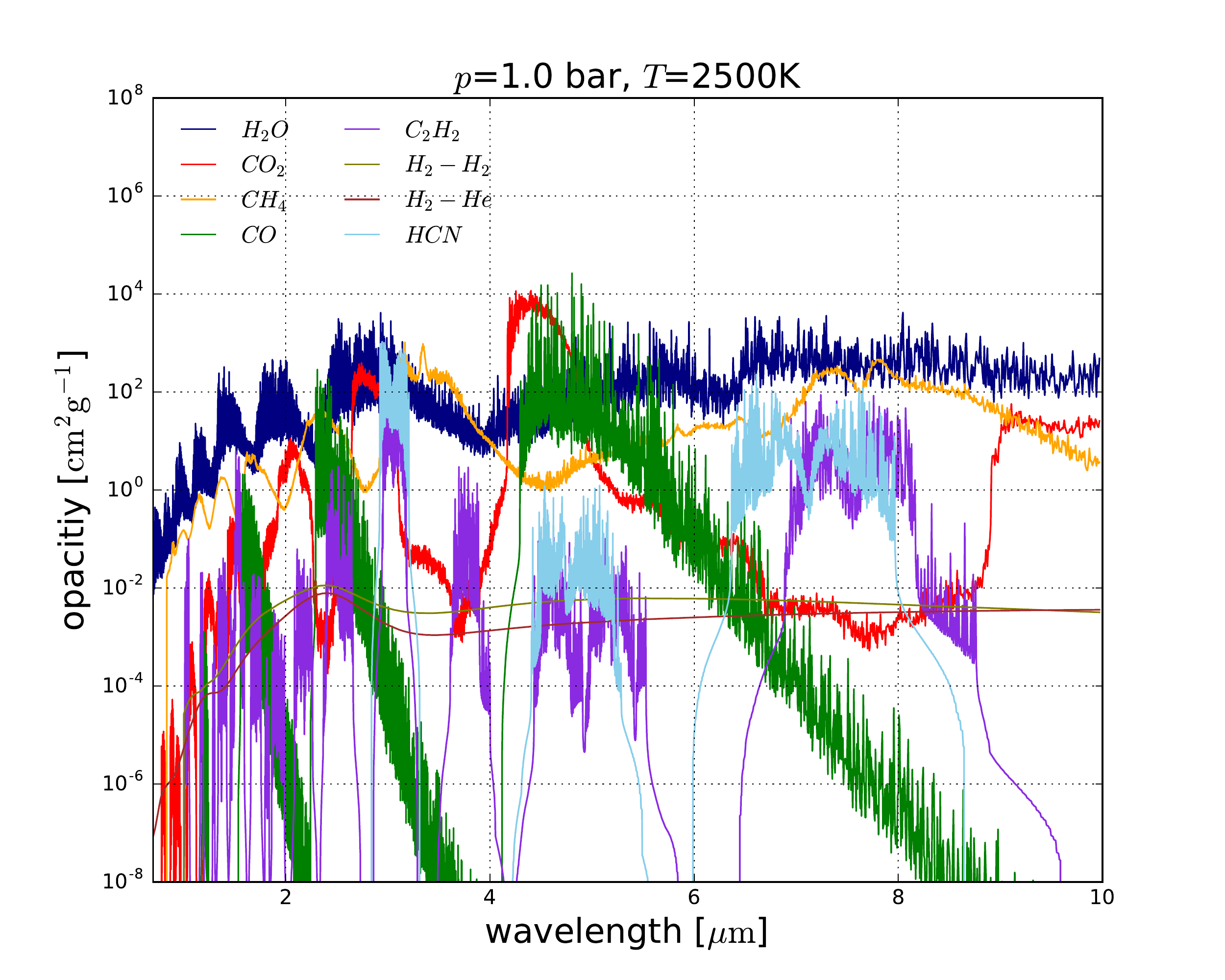}
\end{center}
\vspace{-0.1in}
\caption{Examples of opacities computed using our \texttt{HELIOS-K} opacity calculator \citep{gh15} for a temperature of 2500 K and a pressure of 1 bar.  The \texttt{ExoMol} database is the source of our H$_2$O and CH$_4$ opacities.  The CO and CO$_2$ opacities are from \texttt{HITEMP}, while the C$_2$H$_2$ and HCN opacities are from \texttt{HITRAN}.}
\label{fig:opacities}
\end{figure}

Our nested-sampling retrieval code, \texttt{HELIOS-R}, and computational setup was previously described in \cite{lavie17}.  The stellar and exoplanetary parameters are taken from \cite{hebb09} and \cite{chan11}.  Our nested-sampling \citep{feroz09} retrievals typically use 8 parallel runs of 4000 live points each.  The model atmosphere is divided into 100 discrete layers.  At every wavelength, the propagation of flux is performed using a direct, analytical solution of the radiative transfer equation in the limit of pure absorption \citep{hml14}.  The opacities are computed using our customized, open-source opacity calculator, \texttt{HELIOS-K} \citep{gh15}, which takes the \texttt{HITEMP} \citep{rothman10}, \texttt{HITRAN} \citep{rothman96,rothman13} and \texttt{ExoMol} \citep{barber06,y13,yt14} spectroscopic databases as inputs to compute the line shapes and strengths.  We include the opacities of CO, CO$_2$, CH$_4$, C$_2$H$_2$, H$_2$O and HCN, as well as collision-induced absorption associated with H$_2$-H$_2$ and H$_2$-He.  Figure \ref{fig:opacities} shows examples of the opacities computed.  We use the opacity sampling method with a spectral resolution of 1 cm$^{-1}$.  Our line-wing cutoff is 100 cm$^{-1}$ applied to all of the spectral lines.  We use the analytical temperature-pressure profiles originally derived by \cite{guillot10}, and later generalized to include scattering by \cite{hhps12} and \cite{hml14}.  These profiles enforce radiative equilibrium (local energy conservation) by construction, but are too isothermal at high altitudes due to the assumption that the Planck, absorption and flux mean opacities are equal.  By numerical experimentation (not shown), we find that the temperature-pressure profile in the limit of pure absorption suffices for our purposes, which is to describe the shape of the profile with as few parameters as possible: $\kappa_{\rm IR}$ (the mean infrared opacity associated with the temperature-pressure profile) and $\gamma$ (the ``greenhouse parameter", which is the ratio of the mean optical/visible to mean infrared opacities).  There is no attempt to seek self-consistency between these parameters and the wavelength-dependent opacities used.\footnote{None of the practitioners of atmospheric retrieval are currently able to do this.}  Atmospheres without and with temperature inversions have $\gamma<1$ and $\gamma>1$, respectively.

\begin{figure}[!h]
\vspace{-0.1in}
\begin{center}
\includegraphics[width=\columnwidth]{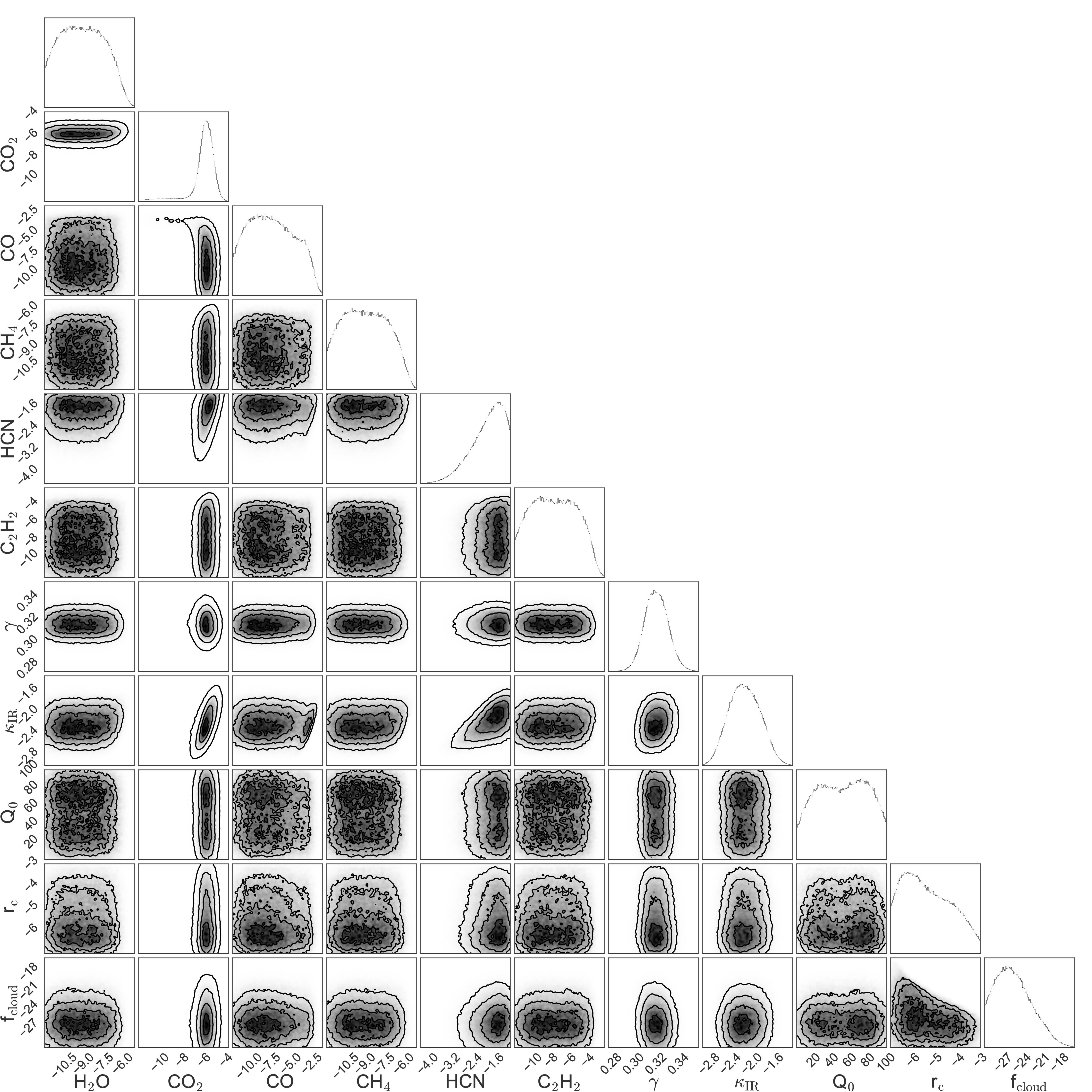}
\includegraphics[width=\columnwidth]{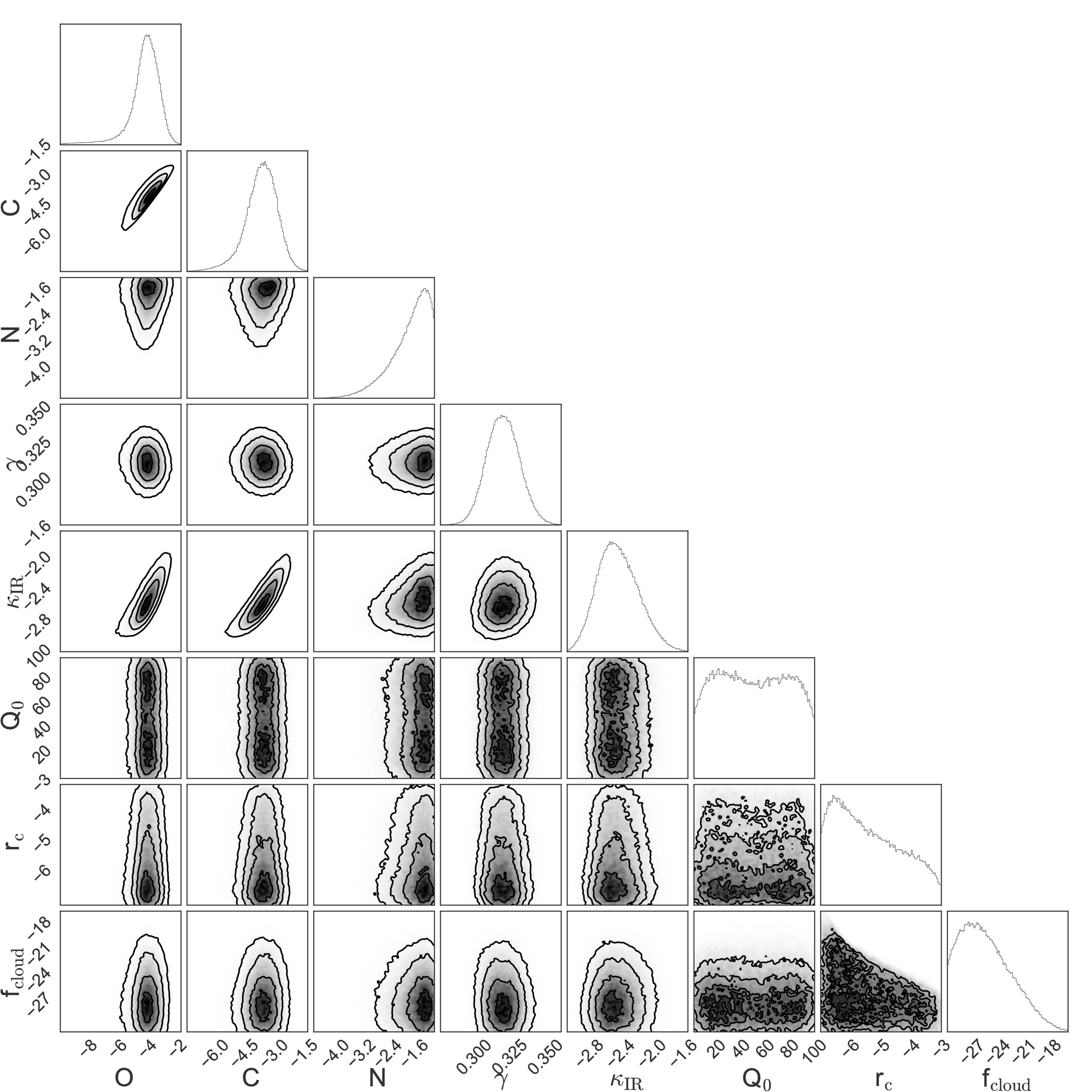}
\end{center}
\vspace{-0.1in}
\caption{Posterior distributions of fitting parameters for cloudy retrieval models with unconstrained chemistry (top panel) and equilibrium chemistry (bottom panel) with log-uniform priors.  $\kappa_{\rm IR}$ has physical units of m$^2$ kg$^{-1}$, while $r_{\rm c}$ is given in m.  The rest of the parameters are dimensionless.}
\label{fig:posterior}
\vspace{-0.1in}
\end{figure}

For chemistry, we consider two types of models: unconstrained and equilibrium chemistry.  The former is the typical approach, which assumes mixing ratios that are constant throughout the atmosphere and uses them as fitting parameters.  In other words, no chemistry is actually being considered.  The latter enforces chemical equilibrium via the analytical formulae of \cite{ht16}, who validated these formulae against calculations of Gibbs free energy minimization and demonstrated that they are accurate at the $\sim 1\%$ level or better.  For chemical-equilibrium models, the fitting parameters are C/H, O/H and N/H.  The prior distribution of C/O is roughly uniform, unlike for unconstrained chemistry where it is double-peaked \citep{line13}.  In chemical equilibrium, specifying the elemental abundances allows all of the molecular abundances to be computed, with no parametric freedom, given a temperature and pressure.

We are agnostic about the terms ``cloud" and ``haze" and use them interchangeably for this study.\footnote{These terms are either used to distinguish between size (Earth science convention) or formation origin (planetary science convention), and there is no consensus within the exoplanet community on their usage.}  We implement the simplified cloud model introduced by \cite{lee13} and used by \cite{lavie17}, which describes a monodisperse population of spherical cloud particles with radii $r_{\rm c}$, cloud volume mixing ratio $f_{\rm cloud}$ and a single composition (represented by the parameter $Q_0$).  Refractory and volatile cloud species have $Q_0 \sim 1$ and $\sim 10$, respectively.  This cloud model accommodates both small and large particles, and correctly reproduces the limits of Rayleigh and grey scattering.  It is based on the notion that curves of the extinction coefficient have a roughly universal shape \citep{pierrehumbert}.

\section{Results}

\begin{figure*}
\vspace{-0.1in}
\begin{center}
\includegraphics[width=0.8\columnwidth]{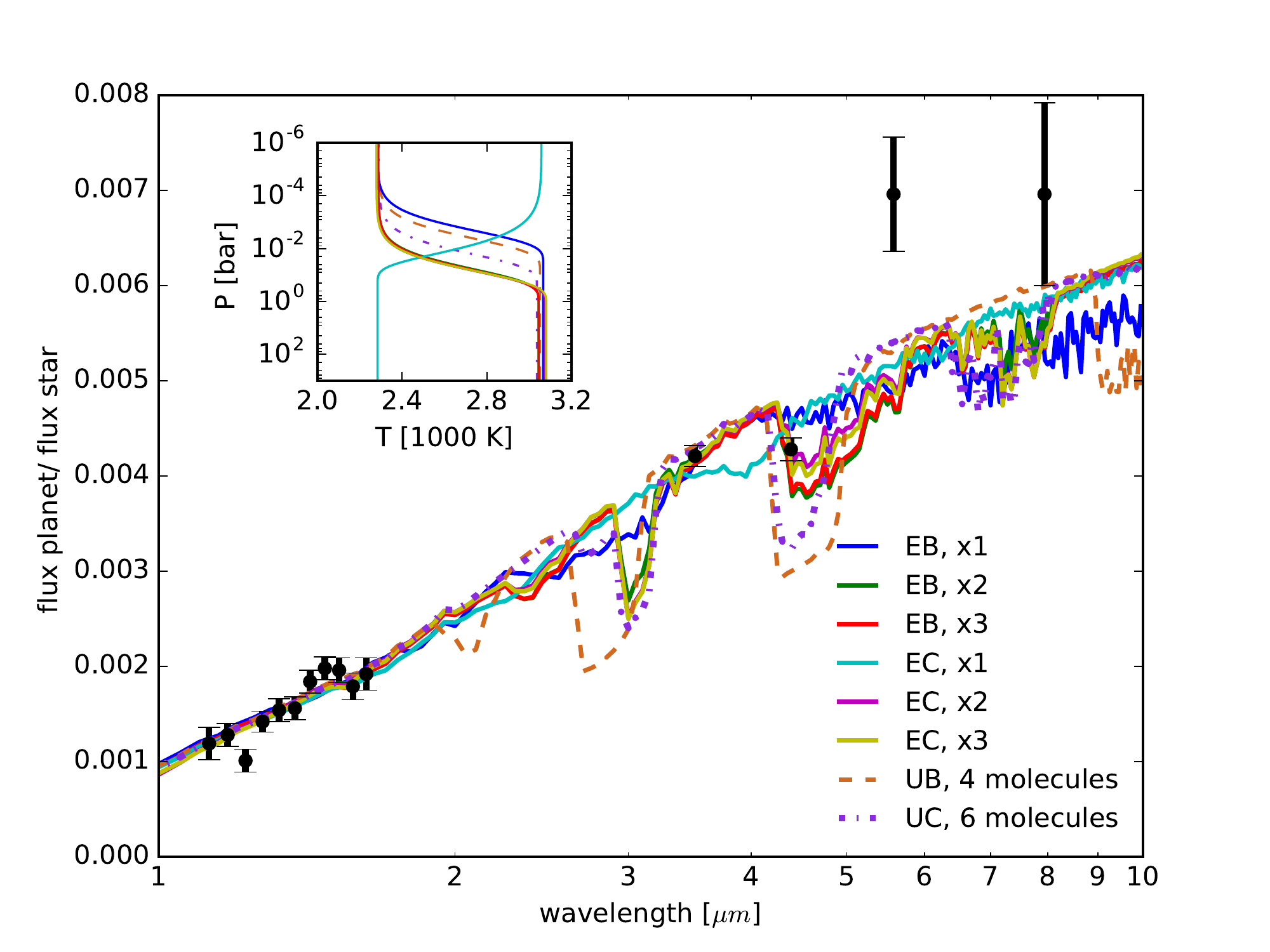} 
\includegraphics[width=0.8\columnwidth]{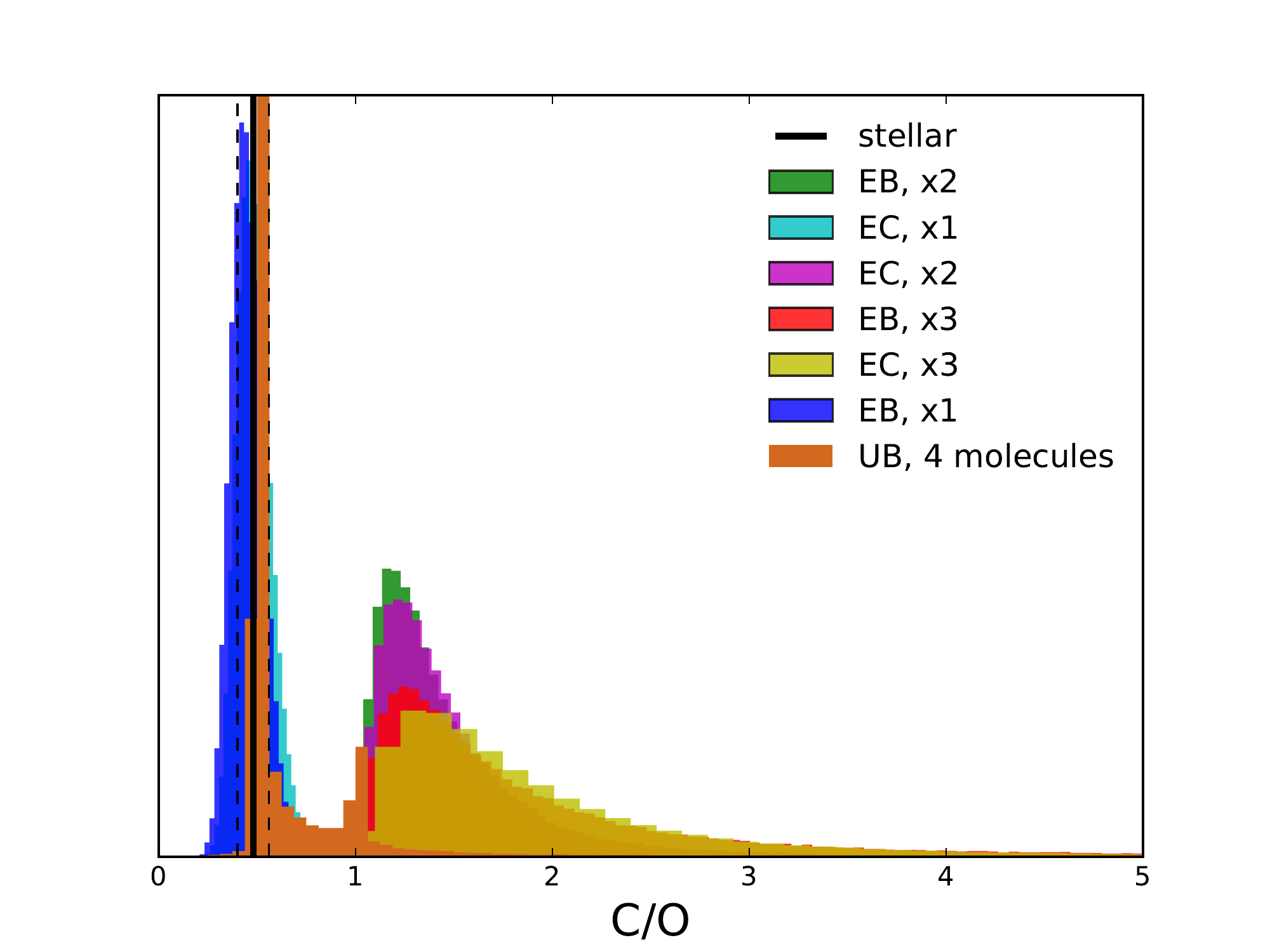}
\includegraphics[width=0.8\columnwidth]{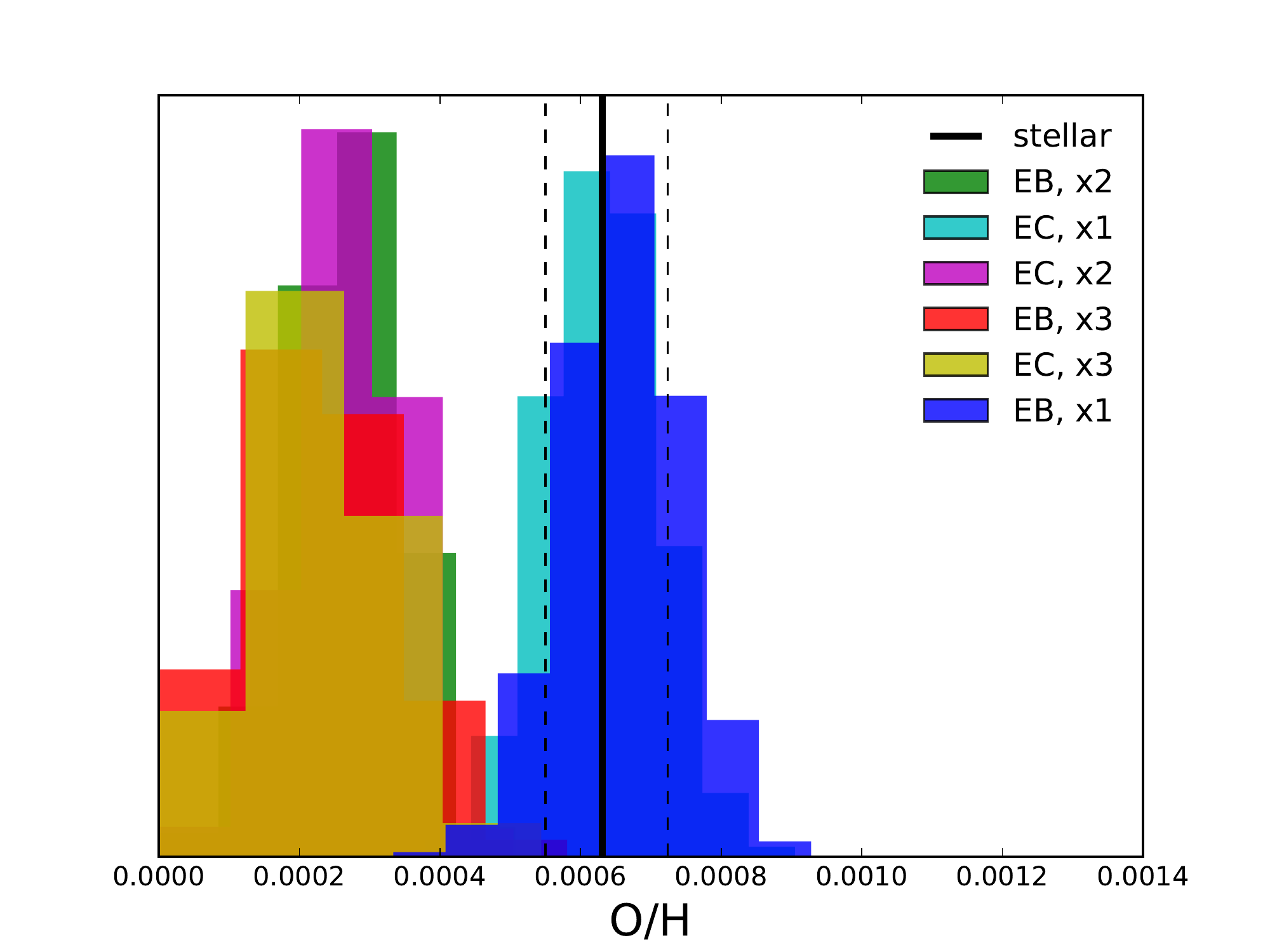}
\includegraphics[width=0.8\columnwidth]{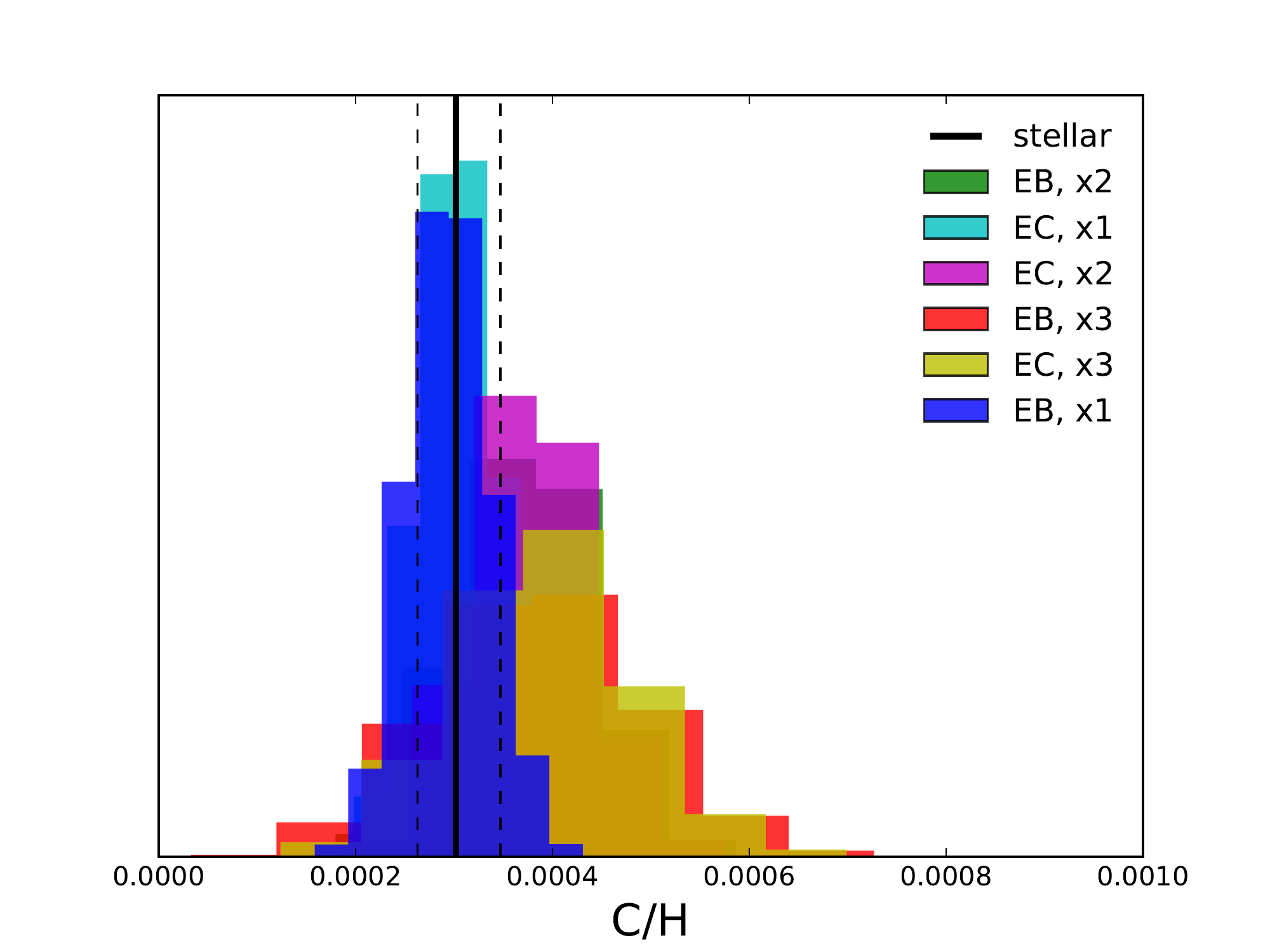}
\includegraphics[width=0.8\columnwidth]{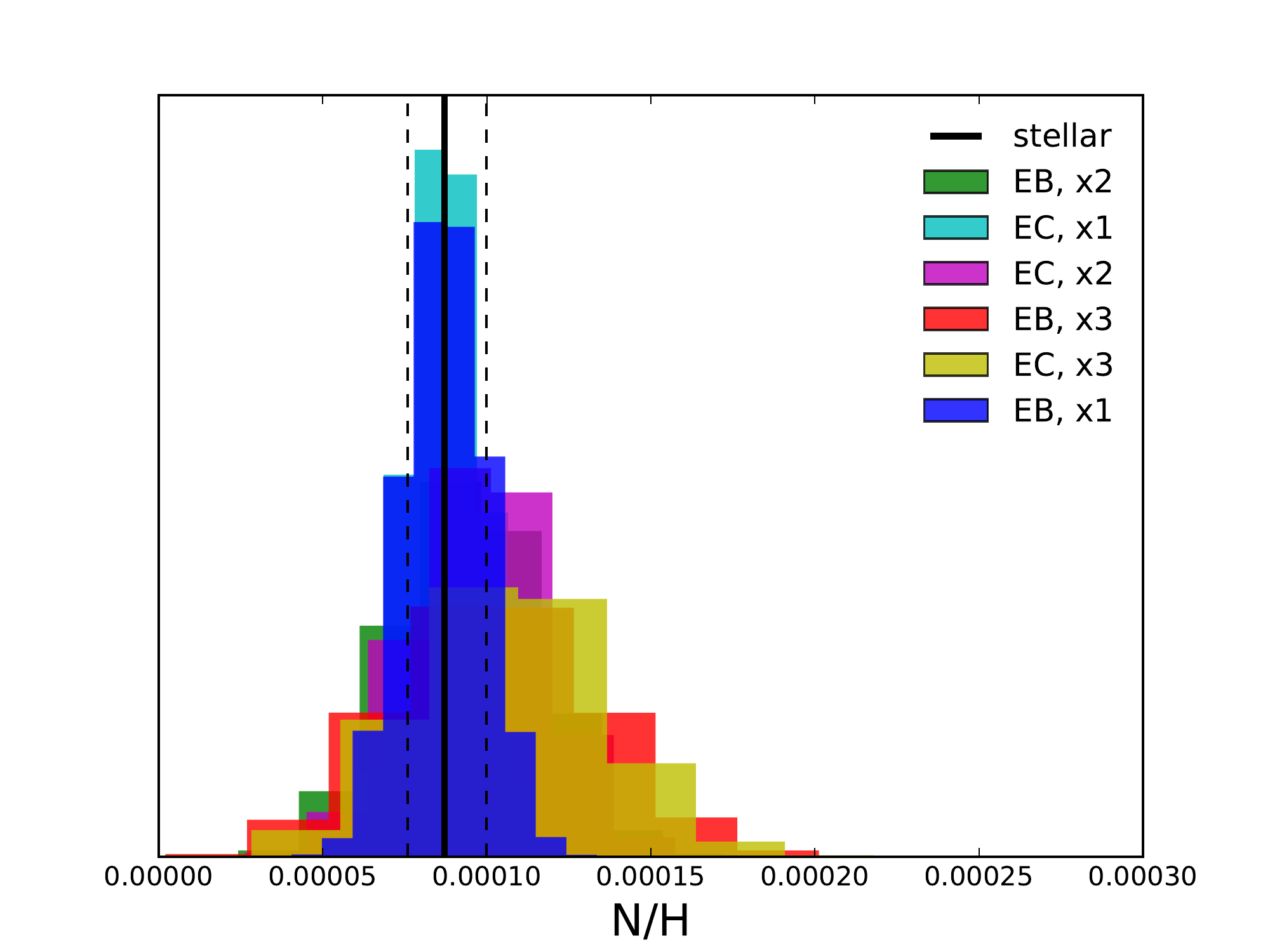}
\includegraphics[width=0.8\columnwidth]{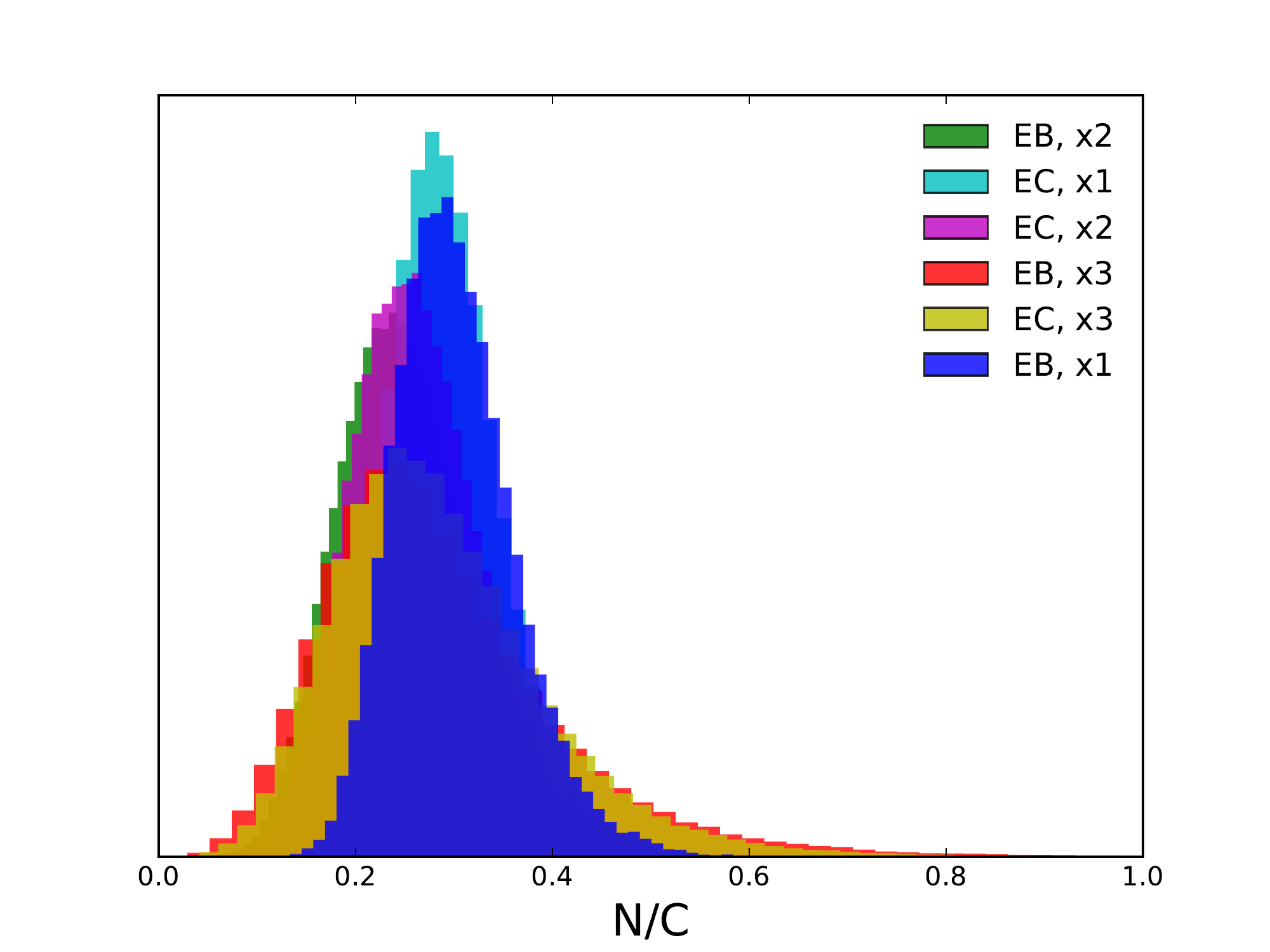}
\end{center}
\vspace{-0.1in}
\caption{Best-fit spectra (top-left panel) and the posterior distributions of C/O (top-right panel), O/H (middle-left panel), C/H (middle-right panel), N/H (bottom-left panel) and N/C (bottom-right panel) for the 6 retrieval models with equilibrium chemistry.  The cloudfree and cloudy models are labeled ``EB" and ``EC", respectively.  The models labeled ``$\times 1$", ``$\times 2$" and ``$\times 3$" adopt Gaussian widths on the prior distributions of the elemental abundances that are once, twice and thrice the measurement errors of the stellar elemental abundances, respectively.  The model labeled ``UB, 4 molecules" assumes unconstrained chemistry and a cloudfree atmosphere with CO, CO$_2$, H$_2$O and CH$_4$ only, and is included as a reference to models previously published in the literature.  The model labeled ``UC, 6 molecules" assumes unconstrained chemistry and a cloudy atmosphere, and is included for completeness as it gives an unrealistic/unphysical abundance for HCN.  The marginal posterior distributions are all normalized to have unity area.  Note that the $\mbox{C/O} \approx 0.5$ peak for the ``UB, 4 molecules" model extends beyond the plot and we have truncated it for clarity.}
\label{fig:co_spectra}
\end{figure*}

\begin{figure}
\begin{center}
\includegraphics[width=0.75\columnwidth]{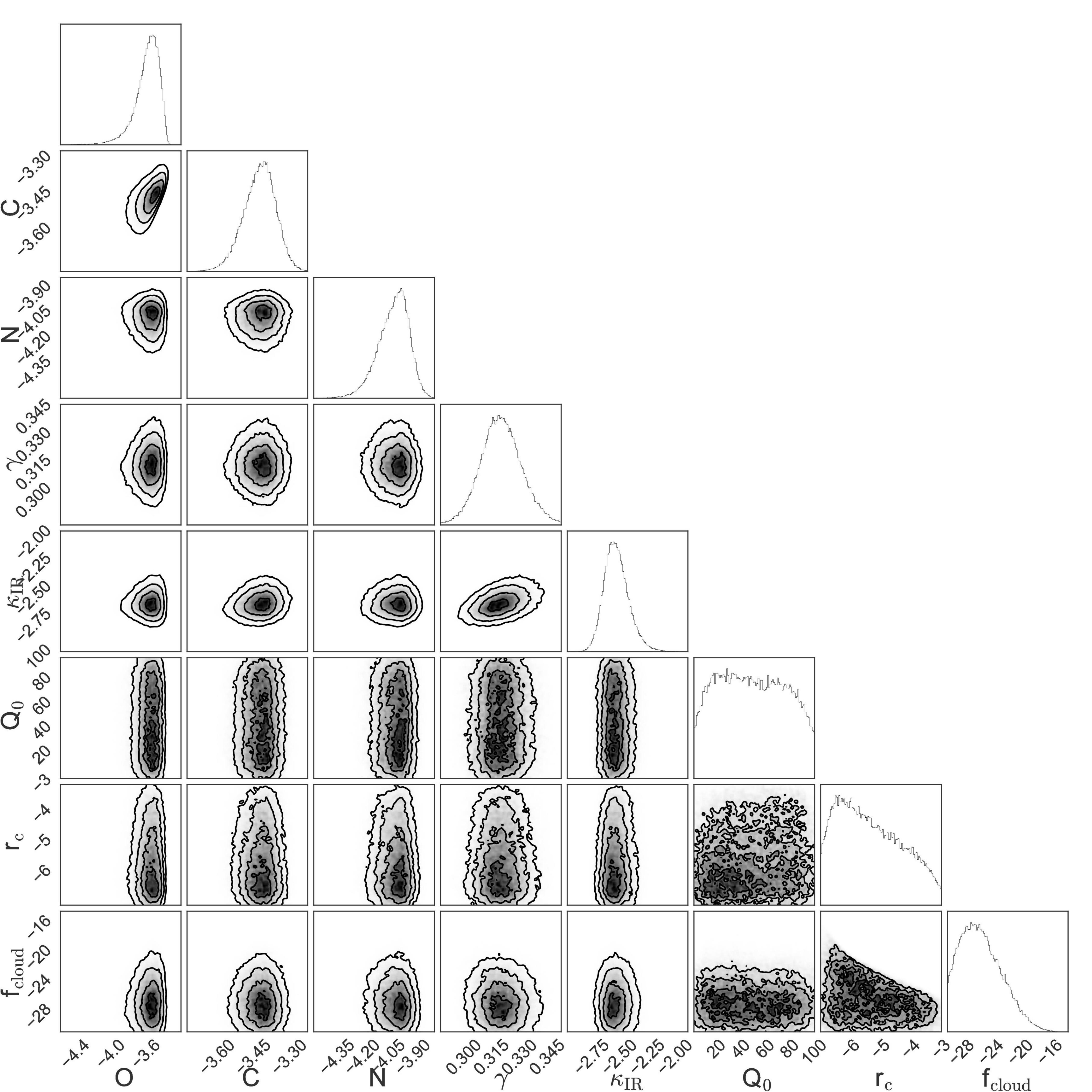} 
\includegraphics[width=0.75\columnwidth]{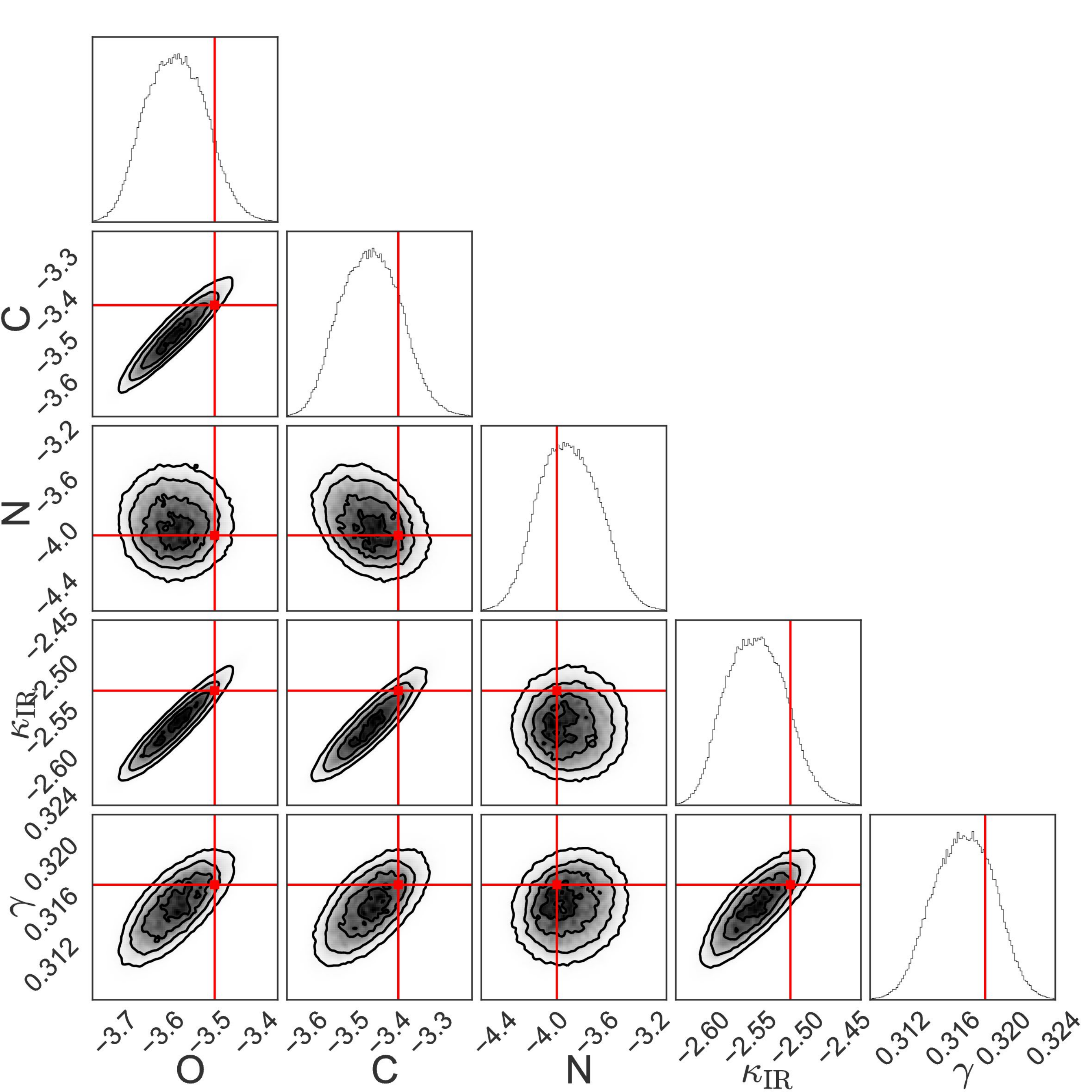} 
\includegraphics[width=0.75\columnwidth]{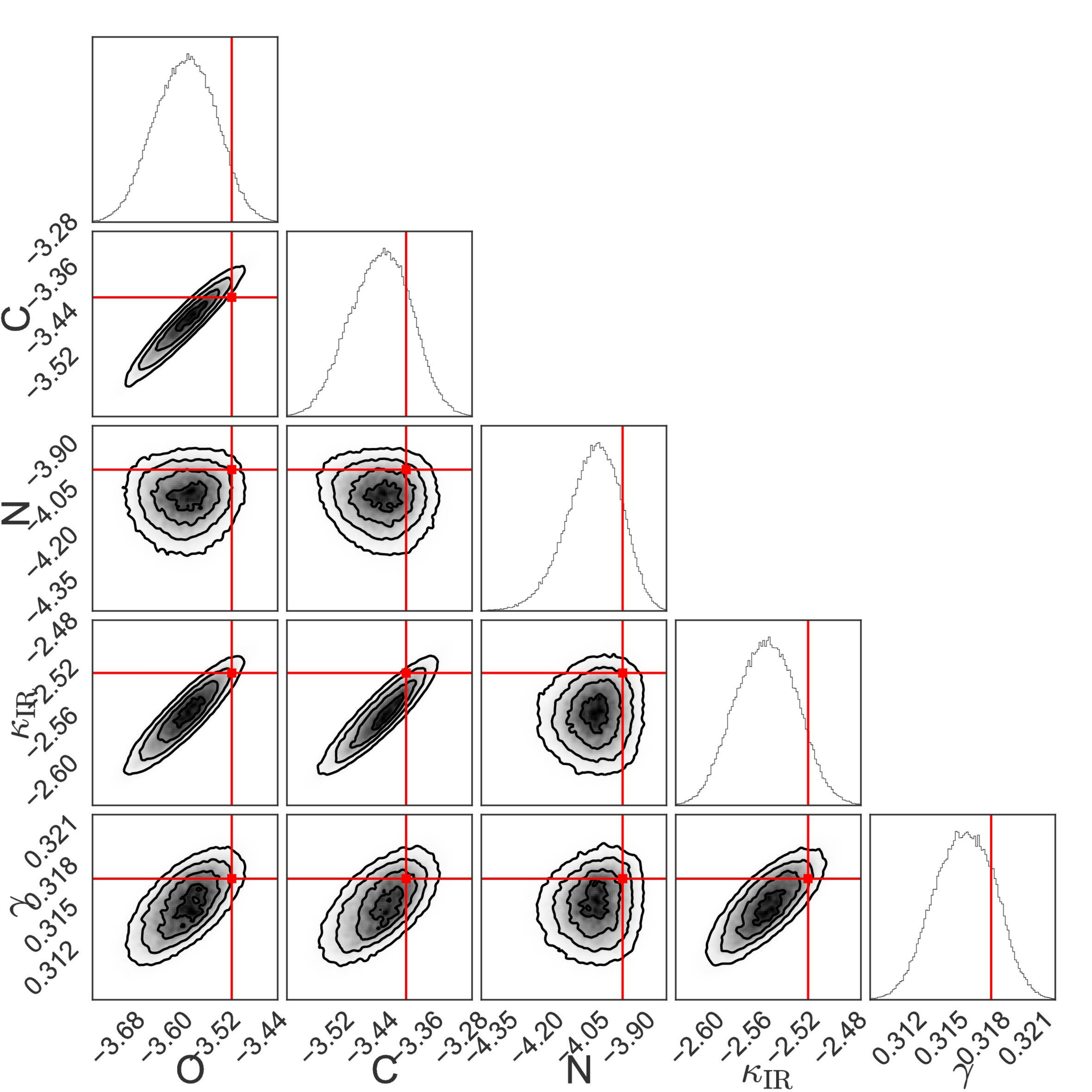} 
\end{center}
\vspace{-0.1in}
\caption{Posterior distributions of fitting parameters for the equilibrium-chemistry model.  $\kappa_{\rm IR}$ has physical units of m$^2$ kg$^{-1}$, while $r_{\rm c}$ is given in m.  The rest of the parameters are dimensionless.  Top panel: Retrieval on WFC3 and Spitzer data with a Gaussian width on the prior distribution of the elemental abundances that is twice the measurement error of the stellar elemental abundances (i.e., the ``EC,$\times 2$" model).  Middle panel: Retrieval on mock JWST data (see text for more details) for EB,$\times2$ cloudfree model.  Bottom panel: Cloudfree retrieval on mock JWST data with log-uniform priors.  For the retrievals on mock data, the input values of parameters are given by the straight lines.}
\label{fig:posterior2}
\end{figure}

We begin by presenting a pair of retrieval models that make the common assumption of log-uniform priors (Figure \ref{fig:posterior}).  For the measured emission spectrum of WASP-12b, we use the published data of \cite{stevenson14b} as stated in their Table 3.  For the model with unconstrained chemistry, we set log-uniform priors on the 6 mixing ratios.  For the model with equilibrium chemistry, we set log-uniform priors on the elemental abundances of carbon, oxygen and nitrogen.  As an improvement over the work of \cite{stevenson14b}, we include clouds in our analysis as part of the retrieval (i.e., the cloud parameters are not fixed to preset values).  The other parameters also have log-uniform priors, except for $Q_0$, which has a (linearly) uniform prior.  

In Figure \ref{fig:posterior}, the first thing to notice is that the cloud parameters display degeneracies that match our physical intuition: the mixing ratios are degenerate with cloud composition, particle radius and number density.  In particular, it is possible to set bounds on the cloud particle radius, but the cloud composition is essentially---and unsurprisingly---unconstrained.  When we include only CO, CO$_2$, CH$_4$ and H$_2$O in the retrieval, we reproduce the result of \cite{line14} and \cite{stevenson14b} that unrealistically high abundances for CO$_2$ are obtained (not shown).  (\citealt{hl16} have previously elucidated this implausibility using validated analytical formulae.)  Such high abundances of CO$_2$ drive the retrieval predominantly towards a solution with $\mbox{C/O} \approx 0.5$.  Furthermore, the prior distribution of C/O is double-peaked at 0.5 and 1 \citep{line13}, which appears in the posterior distribution as well (Figure \ref{fig:co_spectra}).

When C$_2$H$_2$ and HCN are included, we obtain the mixing ratio of HCN to be $\sim 10^{-2}$--$10^{-1}$ (top panel of Figure \ref{fig:posterior}).  This is chemically implausible, as suggested by the detailed chemical kinetics calculations of \cite{moses13}, who estimated an upper limit to the mixing ratio of HCN of $\sim 10^{-3}$ for $\mbox{C/O}<2$ and thrice the solar metallicity.  When chemical equilibrium is enforced with log-uniform priors, we obtain N/H$\sim 10^{-2}$--$10^{-1}$ (bottom panel of Figure \ref{fig:posterior}), which is similarly implausible.  These anomalies arise because the opacity of HCN is driving the fit at the wavelengths of the Spitzer photometry \citep{stevenson14b} (Figure \ref{fig:opacities}).  The lesson learned is that the ``simplest" assumption made on the prior distributions of fitting parameters may not be the best one \citep{trotta08}.  Rather, we need to be guided by physics and chemistry.

Motivated by the calculations in Figure \ref{fig:chemistry}, we enforce chemical equilibrium as a prior.  Instead of log-uniform priors, we now set Gaussian priors on the elemental abundances, based on the measured\footnote{Since these priors are based on measurements, they could alternatively be considered as being part of the likelihood.} WASP-12 values by \cite{teske14}: C/H$_\star=3.02^{+0.45}_{-0.39} \times 10^{-4}$ and O/H$_\star=6.31^{+0.93}_{-0.81} \times 10^{-4}$.  Since \cite{teske14} did not report measured N/H values, we use N/O$_\odot=0.138$ \citep{lodders03} to transform O/H into N/H$_\star=8.71^{+1.28}_{-1.12} \times 10^{-5}$.  We additionally compute models with Gaussian widths that are twice and thrice the measurement errors.  The top panel of Figure \ref{fig:co_spectra} shows that the cloudfree model with $\times 1$ the measurement error as the Gaussian width produces a posterior distribution of C/O that is unsurprisingly peaked at the measured C/O$_\star=0.48$ value of WASP-12.  In other words, we simply reproduce the (tight) prior.  Of greater interest are the posterior distributions when the widths of the Gaussian priors are doubled or tripled, which peak just above a C/O value of unity and trail off as it becomes 2--3.  This outcome of a carbon-rich dayside atmosphere of WASP-12b is independent of whether clouds are included in the analysis, because the cloud layer is optically thin.  The posterior distribution of O/H is substellar, while that of C/H is slightly superstellar but still consistent with being stellar.  Our posterior distributions for C/H, O/H and C/O are broadly consistent with those reported by \cite{madhu11,madhu14a} and \cite{madhu14b}.  We note that increasing the Gaussian widths of the priors to 8 times the measurement errors does not alter our qualitative conclusions (not shown).


Another surprising outcome of this set of 6 retrievals is the shape of the temperature-pressure profile (bottom panel of Figure \ref{fig:co_spectra}).  While the best-fit spectra look similar among the 6 different cases, the temperature-pressure profile for the cloudy $\times 1$ model exhibits a temperature inversion that is entirely driven by the retrieval attempting to fit the four Spitzer photometric points.  When the Gaussian width on the priors is doubled or tripled, the temperature inversion disappears.  For illustration, the top panel of Figure \ref{fig:posterior2} shows the posterior distributions for the cloudy case with $\times 2$ the measurement errors for the Gaussian width of the priors.

\section{Discussion}


\subsection{Implications for formation and comparison to previous studies}

Generally, it is challenging to make a hot Jupiter with substellar O/H \citep{brew17}.  Several studies have previously explored the link between the formation and migration history of hot Jupiters and their atmospheric chemistry.  \cite{madhu14b} predicted that the formation of gas-giant exoplanets at large orbital distances via gravitational instability, from a solar-composition protoplanetary disk, and their subsequent migration inwards via disk-free mechanisms produces hot Jupiters with stellar C/H, substellar O/H and superstellar C/O.  Our retrieval outcomes are consistent with this scenario.  If the disk is instead constructed with molecular abundances based on observations of ice and gas in protoplanetary disks \citep{oberg11}, then it produces hot Jupiters with C/H and O/H that are both substellar.  Core accretion with disk-free migration produces C/H and O/H that are either both substellar, both stellar or both superstellar---neither of these scenarios are consistent with our retrieval outcomes.  Core accretion with disk migration produces superstellar values for both C/H and O/H.  

An active topic of debate concerns the role of pebbles in the protoplanetary disk \citep{ok10,lj12}.  Pebbles are intermediate-sized solids with Stokes numbers on the order of unity, which are imperfectly coupled to the disk gas; their exact sizes are a function of the local conditions of the disk.  The drift of pebbles across the CO$_2$, CO and H$_2$O snowlines is capable of locally altering the values of C/H, O/H and C/O in a disk \citep{oberg16}.  The key difference between pebbles and regular planetesimals is that, to zeroth order, pebbles are purportedly able to accrete onto the core of the exoplanet directly without polluting the atmosphere, implying that the elemental abundances range from being substellar to stellar.  In the scenario depicted by \cite{madhu17}, hot Jupiters accrete most of their gas within the H$_2$O snowline \citep{ad14}, which naturally yields a stellar C/H, substellar O/H and superstellar $\mbox{C/O} \approx 0.7$--0.8.  At face value, this is at odds with our finding that $\mbox{C/O} \approx 1$--2.  Any erosion of the core tends to drive C/H and O/H to superstellar values and C/O to substellar values, further increasing the discrepancy between the theoretical prediction and our inferred posterior distributions.  An alternative scenario is that WASP-12b formed at large orbital distances (as a cold Jupiter) via pebble accretion and migrated inwards via a disk-free mechanism.  In such a scenario, \cite{madhu17} predict $\mbox{O/H} \approx 0.2$--$0.5 ~\mbox{O/H}_\star$, $\mbox{C/H} \approx 0.5$--$0.9 ~\mbox{C/H}_\star$ and $\mbox{C/O} \approx 1$.  Our retrieved posterior distributions are consistent with such a scenario.  Based on the inferred substellar O/H and superstellar C/O values, \cite{brew17} claimed another hot Jupiter, HD 209458b, to also have undergone disk-free migration.

We note that WASP-12b is part of a triple star system \citep{bechter14} and has a measured spin-orbit alignment of $59^{+15}_{-20}$ degrees, which may be consistent with the disk-free migration scenario.

Our retrieved posterior distributions are inconsistent with the in-situ formation of WASP-12b \citep{batygin16,boley16}, which \cite{madhu17} predict to yield $\mbox{O/H} \approx 0.8$--$1.5 ~\mbox{O/H}_\star$, $\mbox{C/H} \approx \mbox{C/H}_\star$ and $\mbox{C/O} \approx 0.4$--0.7.  \cite{ad17} suggests that to produce $\mbox{C/O} \gtrsim 1$ via in-situ formation requires that the parent star has $\mbox{C/O} \approx 0.8$.

Generally, our finding of substellar values for O/H provides counter-evidence against late-time planetesimal accretion or core erosion.  Both processes would enrich the atmosphere of WASP-12b to beyond its stellar values.  Furthermore, the posterior distribution of N/C, which is consistent with being solar \citep{lodders03}, provides clues on the original site of formation in the outer protoplanetary disk \citep{oberg16}.

\subsection{Are retrievals of JWST data in the prior-dominated regime?}

Our findings beg the question: are retrievals of James Webb Space Telescope (JWST) spectra also in the prior-dominated regime?  To address it specifically for WASP-12b, we produce mock spectra with a resolution of 100 over the wavelength range of 0.7 to 5 $\mu$m.  We assume measurement uncertainties of 100 ppm.  The middle and bottom panels of Figure \ref{fig:posterior2} show the posterior distributions of parameters from retrievals assuming log-uniform and Gaussian priors, respectively.  In both cases, the retrieved parameter values are essentially the same and within $\sim 30\%$ of the true (input) values, suggesting that the interpretation of JWST spectra will not be in the prior-dominated regime.

\acknowledgments
We acknowledge partial financial support from the Center for Space and Habitability (CSH), the PlanetS National Center of Competence in Research (NCCR), the Swiss National Science Foundation and the Swiss-based MERAC Foundation.


\end{document}